\documentclass[useAMS,usenatbib]{mn2e}

\newcommand{\be} {\begin{equation}}

\newcommand{\ee} {\end{equation}}

\newcommand{\xmm}{{\em XMM--Newton}}
\newcommand{\XMM}{{\em XMM--Newton}}

\newcommand{\RXTE}{{\em RXTE}}
\newcommand{\swift}{{\em Swift}}

\newcommand{\bc}{\begin{center}}
\newcommand{\ec}{\end{center}}
\def\ltsima{$\; \buildrel < \over \sim \;$}
\def\lsim{\lower.5ex\hbox{\ltsima}}
\def\loe{\lower.5ex\hbox{\ltsima}}
\def\gtsima{$\; \buildrel > \over \sim \;$}
\def\gsim{\lower.5ex\hbox{\gtsima}}
\def\goe{\lower.5ex\hbox{\gtsima}}

\def\ergs{erg\,s$^{-1}$}
\def\ergscm2{erg\,s$^{-1}$cm$^{-2}$}
\def\ergsscm2{erg\,s$^{-2}$cm$^{-2}$}
\def\cm2{cm$^{-2}$}
\def\ss{s\,s$^{-1}$}

\def\ee{1E\,1048.1--5937\,}

\def\1e{1E\,1547.0-5408\,}

\def\sgrnew{SGR\,0501+4516\,}
\def\aj{\ref@jnl{AJ}}                   
             
\def\apj{{ApJ}}                 
\def\apjl{{ApJ}}

\def\aap{{A\&A}}                
\def\aapr{{A\&A~Rev.}}

\def\mnras{{MNRAS}}

\def\nat{{Nature}}

\def\gca{{Geochim.~Cosmochim.~Acta}}

\title[Quiescent state and outburst evolution of SGR\,0501+4516 ]{Quiescent state and outburst evolution of SGR\,0501+4516}
\author[A. Camero, et al.]{A. Camero$^{1}$\thanks{E-mail: camero@ice.cat}, A. Papitto$^{1}$, N. Rea$^{1,2}$, D. Vigan\`o$^{1,3}$, J. A. Pons$^3$, A. Tiengo$^{4,5,6}$,\newauthor S. Mereghetti$^6$, R. Turolla$^{7,8}$, P. Esposito$^{6,4}$, S. Zane$^7$, G.L. Israel$^9$, D. G\"{o}tz$^{10}$\vspace*{0.2cm}\\ 
 $^{1}$ Institut de Ci\`{e}ncies de l'Espai, (IEEC-CSIC), Campus UAB, Fac. de Ci\`{e}ncies, Torre C5, parell, 2a planta, 08193 Barcelona, Spain\\
 $^2$ Astronomical Institute 'Anton Pannekoek', University of Amsterdam, Science Park 904, Postbus 94249, 1090 GE, Amsterdam, the\\ Netherlands\\
 $^3$ Departament de F\'{i}sica Aplicada, Universitat d'Alacant, Ap. Correus 99, 03080, Alacant, Spain\\
 $^4$ INFN -- Istituto Nazionale di Fisica Nucleare, Sezione di Pavia, via A.~Bassi 6, 27100 Pavia, Italy \\
 $^5$ IUSS --  Istituto Universitario di Studi Superiori, piazza della Vittoria 15, I-27100 Pavia, Italy\\
 $^6$ INAF -- Istituto di Astrofisica Spaziale e Fisica Cosmica, via E.~Bassini 15, I-20133, Mlano, Italy \\
 $^7$ Mullard Space Science Laboratory, University College London, Holmbury St. Mary, Dorking, Surrey, RH5 6NT, UK\\ 
 $^8$ Universit\`a di Padova, Dipartimento di Fisica, via Marzolo 8, I-35131 Padova, Italy\\ 
 $^9$ INAF -- Osservatorio Astronomico di Roma, via Frascati 33, 00040, Monte Porzio Catone (RM), Italy\\
 $^{10}$ AIM (UMR 7158 CEA/DSM-CNRS-Universit\'e Paris Diderot) Irfu/Service d'Astrophysique, Saclay, F-91191 Gif-sur-Yvette Cedex, \\ France\\
}

\input psfig.sty

\begin{document}

\pagerange{\pageref{firstpage}--\pageref{lastpage}} \pubyear{2013}

\maketitle

\label{firstpage}

\begin{abstract}

We report on the quiescent state of the Soft Gamma Repeater SGR\,0501+4516 observed by \XMM\,  on 2009 August 30. The source exhibits an absorbed  flux $\sim$75 times lower than that measured at the peak of the 2008 outburst, and a rather soft  spectrum, with the same  value of the blackbody temperature  observed with \textit{ROSAT} back in 1992.  This new observation is put into the context of all existing X-ray data since its discovery in August  2008,  allowing us to complete the study of the timing and spectral evolution of the source from outburst until its quiescent state. The set of deep \XMM\, observations performed during the few-years timescale of its outburst allows us to monitor the spectral characteristics of this magnetar as a function of its rotational period, and their evolution along these years. After the first $\sim10$ days, the initially hot and bright surface spot progressively cooled down during the decay. We discuss the behaviour of this magnetar in the context of its simulated secular evolution, inferring a plausible dipolar field at birth of $3\times10^{14}$\,G, and a current (magneto-thermal) age of $\sim10$\, kyr. 

\end{abstract}


%%%%%%%%%%%%%%%%%%%%   TABLE: OBSERVATIONS %%%%%%%%%%%%%%%%%%%%%%%%%%%%%%%%%%%%%%%%%%
\begin{table*}
\begin{center}
\caption{Summary of all the available \XMM\, observations of \sgrnew\, since the discovery outburst. The exposure time refers to the pn camera. Count-rates  are background-corrected. The pulsed fraction is defined as
  the background-subtracted $(max-min)/(max+min)$ in the 0.3-12\,keV
  energy band.\label{obslog}}
\begin{tabular}{lcccccccc}
\hline
\hline
\multicolumn{4}{c}{}& \multicolumn{2}{c}{}\\
\hline
 \multicolumn{1}{l}{Parameters}  & \multicolumn{1}{c}{2008-08-23} &
 \multicolumn{1}{c}{2008-08-29} & \multicolumn{1}{c}{2008-08-31} &
 \multicolumn{1}{c}{2008-09-02} & \multicolumn{1}{c}{2008-09-30} &
 \multicolumn{1}{c}{2009-08-30}\\
\hline
Start (UT)  & 01:07:36  & 07:10:28  &  12:09:45 &  10:00:38 & 02:18:44 & 14:45:41\\
End (UT)   & 14:35:33 & 13:58:20  &  14:59:58  & 15:41:49 & 11:22:15 & 05:47:52 \\
Exposure (ks)   &  48.9 & 24.9 &  10.2 &  20.5 & 31.0 & 37.8\\
Counts/s    & $8.520(16)$  &  $7.08(2)$ &  $6.60(3)$  & $6.05(2)$  &  $3.23(1)$ & $0.769(5)$\\
\hline
P. Period (s)       & 5.7620694(1) & 5.7620730(1) & 5.7620742(1) & 5.7620754(1)   & 5.7620917(1)  & 5.7622571(2)\\
P. Fraction (\%)    &  41(1) & 35(1)  & 38(1) & 38(1) & 43(1)  & 45(6)\\
N. bursts    & 80 & 2  &  0  & 0  & 0  & 0\\
\hline
\hline
\end{tabular}

\end{center}
\end{table*}

%%%%%%%%%%%%%%%%%%%%%%%%%%%%%%%%%%%%%%%%%%%%%%%%%%%%%%%%%%%%%%%%%%%%%%%%%%

%%%%%%%%%%%%%%%%%%%%  OUTBURST EVOL  %%%%%%%%%%%%%%%%%%%%%%%%%
\begin{figure}
\hspace{-0.3cm}\psfig{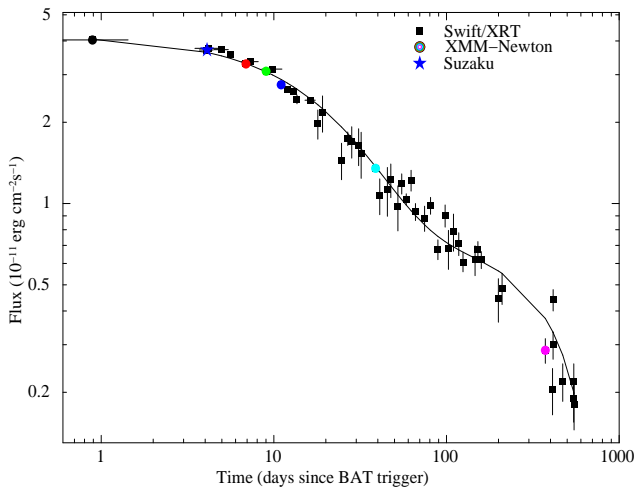}
\caption{The outburst decay of the persistent X-ray flux of \sgrnew\, fitted with a double exponential function (see  text for details). We refer here as BAT trigger: MJD 54700.0 12:41:59.000 (UT). The fluxes are absorbed and in the 1--10\,keV energy range.}
\label{figfluxdecay}
\end{figure}
%%%%%%%%%%%%%%%%%%%%%%%%%%%%%%%%%%%%%%%%%%%%%%%%%%%%%%%%%%%%%%%%

\section{Introduction}

Magnetars are isolated neutron stars with a  persistent X-ray emission  of $L_{\rm X}\sim10^{34}-10^{36}$\ergs, rotating at spin periods of $\sim2-12$\,s and with large period derivatives ($10^{-13}-10^{-10}$\ss) in respect to radio pulsars. Occasionally  these objects also emit bursts and outbursts. Anomalous X-ray Pulsars (AXPs) and Soft Gamma Repeaters (SGRs) are two observational manifestations of magnetars.  The radiation observed from these objects is  believed to be powered by  the decay of their strong magnetic fields ~\citep[see][and Rea \& Esposito 2011 for recent reviews]{Mereghetti08}.

SGRs are characterized by periods of activity during which they emit short bursts in the hard X-ray/soft gamma-ray energy range ($t\sim0.1-0.2$\,s; $L_X\sim10^{38}-10^{41}$ erg/s). In addition, they have been observed to emit intermediate flares, with typical durations of $t\sim1-60$\,s and  luminosities of $L_X\sim10^{41}-10^{43}$ erg/s, and rarely  Giant Flares ($t\sim200-400$\,s, and peak luminosities larger than $\sim4\times10^{44}$ erg/s) \citep[][and references therein]{Mereghetti08}. The frequent short bursts are supposed to be associated either with small cracks in the neutron star crust due to magnetic diffusion or with  tearing instabilities produced by the sudden loss of magnetic equilibrium, while the intermediate flares can arise from heating of the corona by magnetic reconnection in the stellar magnetosphere  \citep[Thompson $\&$ Duncan 1995;][]{lyubarsky02}. The giant flares would be linked to global rearrangements of the magnetic field in the neutron stars magnetosphere and interior (Thompson $\&$ Duncan 1995). 

Magnetar persistent  X-ray emission is usually characterized by a soft ($\sim 0.5$-- 10 keV) X-ray component interpreted as due to thermal photons emitted from the surface which undergo resonant cyclotron scattering onto mildly relativistic electrons flowing in the magnetosphere (Thompson et al. 2002). In some cases, a hard X-ray component extending up to  $\sim200$\,keV has been observed, possibly due to the same up-scattering process but on a population of more energetic electrons (Baring et al. 2007, Beloborodov et al. 2007, Nobili et al. 2008, Beloborodov 2013). The soft X-ray component (on which this paper focuses) is generally fitted with either a blackbody with a temperature kT$\sim$0.3--0.6\,keV and a power-law with $\Gamma\sim$2--4, or two blackbodies  (kT$_{1}\sim$0.3 keV and kT$_{2}\sim$0.7 keV) \citep{Mereghetti08}, while the hard X-ray component is modelled by a power-law spectrum of an index of $\sim$0.5--1.5.

\sgrnew was discovered on 2008 August 22 by \swift-BAT in correspondence to a series of short X-ray bursts and intermediate flares \citep{holland08,Barthelmy08}.  X-ray pulsations at a period of 5.7\,s were observed by \RXTE\, \citep{Gogus08} and an estimate of the surface dipolar magnetic field (at the pole) $B\simeq2\times 10^{14}$ G was obtained from its period and spin-down rate \citep{Woods08, Rea09, Gogus10}.  Detailed studies of \sgrnew\, in the X-rays have been published by Rea et al. (2009), \cite{Gogus10}, and \cite{Kumar10}.  Radio observations did not reveal any emission in the first days after the beginning of the outburst  \citep{Hessels08, Kulkarni08b, Gelfand08}, remaining undetected afterwards. An optical/infrared counterpart has been identified \citep{Tanvir08, Rea08b, Fatkhullin08, Rol08}.   In addition, optical pulsations have been detected for \sgrnew\ \citep{Dhillon11}.

In this paper, we present spectral and timing analysis of  a new \xmm\ observation of \sgrnew~ carried out one year after the onset of the bursting activity which led to its discovery.   In addition, we used  all the existing data from \xmm, \textit{Swift}-XRT, \textit{RXTE}, \textit{Chandra}, and \textit{Suzaku}  to model the flux and the spin period evolution. 

\section{Observations and reduction}
\label{obs}

\subsection{\textit{XMM-Newton}}

The \xmm\, Observatory (Jansen et al. 2001) observed \sgrnew\,again on 2009 August 30 (see Tab.\ref{obslog}) with the EPIC instruments \citep[pn and MOSs;][]{Turner01, Struder01}.%, and the Reflecting Grating Spectrometer \citep[RGS;][]{denHerder01}.

This new observation was processed using SAS version 12.0.1 with the most up to date calibration files (CCF) available at the time the reduction was performed (December 2012). Standard data screening criteria were applied in the extraction of scientific products. Soft proton flares were not present, resulting in the total on-source exposure time of 37.8\,ks (see also Table \ref{obslog}).  For consistency,  source and background events were also re-extracted  for  previous observations (see Table~\ref{obslog}),  using the same version of the standard software. 

Similar to Rea et al. (2009), here we will report only on the pn results (see Table \ref{obslog} for the pn source count rate for this new observation). The observation with the pn camera was set in {\tt Prime Small Window} mode, using thick filter. We extracted source photons from a circular region with 30\arcsec radius, centered at the source position  \citep[RA=75.278167$^\circ$, Dec=+45.276089$^\circ$,][]{Woods08}. The background was obtained from a similar region but far from the source.  Only photons with PATTERN$\leq 4$ were retained. All the photon arrival times have been converted to refer to the barycenter of the Solar System.

\subsection{Other X-ray observations}

To obtain an accurate timing solution we also extracted the events from  observations carried out by  other X-ray missions.  SGR\,0501+4516 was first monitored with the \textit{Swift}/XRT from 2008 August 26 until 2009 April 19, in windowed timing (WT) mode and for a total exposure of 436\,ks  (see  G{\"o}{\u g}{\"u}{\c s} et al. 2010, for details).  Later on, seven new observations were performed in photon counting (PC) mode from 2009 October 7 to 2010 February 21, and for a total exposure time of 38.5 ks.  To extract all the events we  used \texttt{xselect} as part of the HEAsoft package (version 6.13), and then converted all the event arrival times to the solar system barycenter. For this we selected the same source position used for the \textit{XMM-Newton} data set.

%%%%%%%%%%%%%%% TIMING EVOL.   %%%%%%%%%%%%%%%%%%%%%%%%%%%%%%

\begin{figure}
\psfig{figure=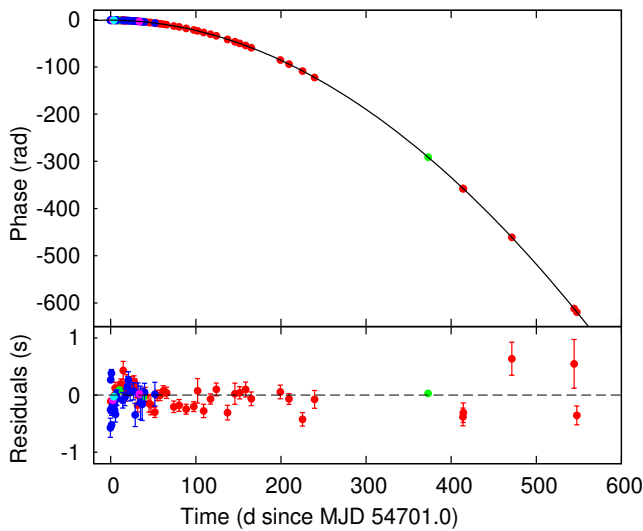,height=7cm,width=8.25cm}
\caption{\textit{Top panel}: The 0.5--10\,keV pulse phase evolution with time (red is \textit{Swift}, blue is \textit{XTE}, green is \textit{XMM}, magenta is \textit{Suzaku}, cyan is \textit{Chandra}). The solid line represents the timing solution. \textit{Bottom panel}: Time residuals with respect to the phase coherent timing solution discussed in the text.}

\label{phasecorr}
\label{phase}
\end{figure}
%%%%%%%%%%%%%%%%%%%%%%%%%%%%%%%%%%%%%%%%%%%%%%%%%%%%%%%%%%%%%%%%

From 2008 August 22 to October 14 the source was monitored with \textit{RXTE}/PCA, with a total exposure time of 82.4 ks in 29 observations (details in  G{\"o}{\u g}{\"u}{\c s} et al. 2010).  %Events  were extracted  and then converted  to the solar system barycenter at the source position. 
The source was also observed with the  Advance CCD Imaging Spectrometer (ACIS) and the High Resolution Camera (HRC) on board \textit{Chandra}, on 2008 August 26 (36.5\,ksec) and September 25 (10\,ksec) (see G{\"o}{\u g}{\"u}{\c s} et al. 2010 for a discussion). %We then extracted source and background  events in the 0.3--10\,keV energy range and applied barycenter correction to all the events. 
\textit{Suzaku} observed SGR\,0501+4516 with the X-ray Imaging Spectrometer (XIS) for $\sim$51\,ks on 2008 August 26--27 \citep[see  ][]{enoto09}.  For all these observations  source and background  events were extracted, and was applied the barycenter correction at the source position.

\section{Timing Analysis and Results}
\label{timing}

%%%%%%%%%%%%%%%%%%%%%%%%    FOLDED PROFILES  AND PF  %%%%%%%%%%%%%%%%%%%%%%%%%%%%%%%%%%%%%%%%%

\begin{center}
\begin{figure}
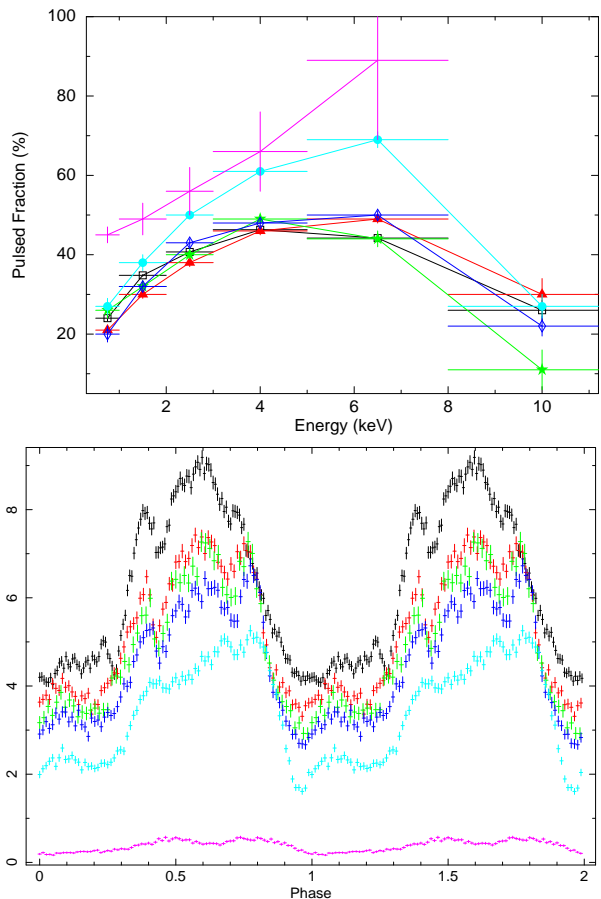

\vbox{
\hspace{-0.75cm}\psfig{figure=fig3_1.ps,width=8.25cm,angle=-90}}
\vspace{0.1cm}
\hspace{-0.75cm}\psfig{figure=fig3_2.ps,width=8.25cm,angle=-90}
\caption{{\em Top panel}: Pulsed fraction
dependence with energy for the the previous 5 \XMM\, observations
 and the one during quiescence. In both panels
the black, red, light green, dark blue, cyan and magenta colors refer to the
six observations ordered by increasing epoch. {\em Bottom panel}: Pulse profiles of the same observations in the same energy band (see also a single profile during quiescence in Fig.7).}
\label{pfenergy}
\end{figure}
\end{center}

%%%%%%%%%%%%%%%%%%%%%%%%%%%%%%%%%%%%%%%%%%%%%%%%%%%%%%%%%%%%%%%%%%%%%%

%%%%%%%%%%%%%%%%%%%%% efolds  %%%%%%%%%%%%%%%%%%%%%%%%%
%\begin{center}
\begin{figure}
\psfig{figure=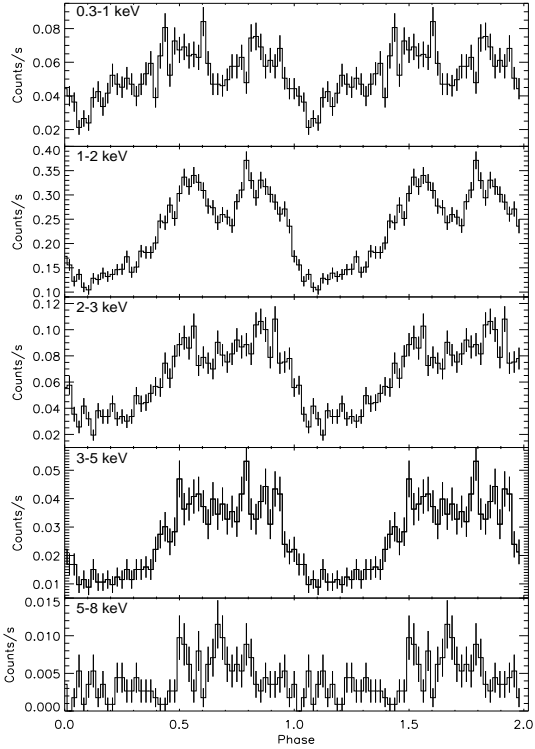,width=7.8cm,height=10cm}
\caption{\textit{XMM-Newton} pulse profiles  as a function of energy  during quiescence state in 2009  for  \sgrnew.}
\label{pulsenergy}
\end{figure}
%\end{center}
%%%%%%%%%%%%%%%%%%%%%%%%%%%%%%%%%%%%%%%%%%%%%%%%%%%%%%%%%%%%%%%%

In order to produce a timing solution for the source covering the time
interval over which data are available (MJD 54701.1--55248.6), we
started from the solution determined by Rea et al. (2009). We
considered data obtained by \textit{Swift}/XRT,
\textit{XMM-Newton}/EPIC pn and \textit{Suzaku}, \textit{RXTE}/PCA and
\textit{Chandra} covering the first $\sim$160 days (interval MJD
54701.1--54859.5). We folded all the available observations around the
best period they determined, $P(T_{0})=5.7620695$ s, where
$T_0=54701.0$ MJD, sampling the profiles so obtained in 12 bins. The
phases of the profiles were obtained from a sinusoidal fit of the
profile. We note that only phases evaluated from profiles detected at
a significance larger than 2$\sigma$ were collected.  On top of a
strong noise component affecting the pulse phases (which leads to a
high value of the fit $\chi^2_r\simeq$ 7.8), we found that a second
derivative significantly improves the modelling with respect to a fit
with a quadratic function (see leftmost column of Table 2), compatible
with the results quoted by Rea et al. (2009).

%%%%%%%%%%%%%%%%%%%  TABLE: TIMING SOLUT.     %%%%%%%%%%%%%%%%%%%%%%%%%%%%%%%%%%%%%%%%%%%%%%

\begin{table}
\centering
%\scriptsize
\caption{Timing solution of SGR\,0501+4516. Quoted errors are evaluated at 1$\sigma$ confidence level, taking into account the number of parameters determined simultaneously, and scaling the formal errors by the rms of residuals, in order to quote the uncertainties which would have been obtained if the errors on the single points were rescaled to give a $\chi^2_r=1$.}          
\label{table1}    
\centering                         
\begin{tabular}{lrr}       
\hline\hline             
Parameter & MJD 54701.1--54859.5 & MJD 54701.1--55248.6  \\
\hline
$T_0$ (MJD) & 54701.0 & 54701.0  \\
$P(T_0)$ (s) & 5.7620699(2) & 5.7620695(1) \\
$\dot{P}(T_0)$ (s/s) & 6.4(1)$\times10^{-12}$ & 5.94(2)$\times10^{-12}$ \\
$<\ddot{P}>$ (s/s$^2$) & -9(3)$\times10^{-20}$ & -1.0(1)$\times10^{-20}$ \\
$\chi^2_r$ &  635.751/81 & 851.511/92 \\
\hline
\hline                            
\end{tabular}
\end{table}

%%%%%%%%%%%%%%%
Finally, to obtain a timing solution for the whole interval ($\sim$
548 days) we added the new observations during quiescent state. We
point out that the uncertainty on the resulting parameters, and in
particular that affecting the second derivative, prevents to
univoquely phase connect the observations obtained by
\textit{XMM-Newton} and \textit{Swift} in quiescence (after MJD
55073.616).  Nevertheless, we tentatively phase connect these
observations, considering the solution giving the lower value of
residuals as the most probable one. The best solution quoted in the
rightmost column of Table 2 has a $\chi^2$ lower by 104, for 92
degrees of freedom, with respect to the second best. However, we
stress that the solution is not unique, since the phase connection is
not warranted by the available data, and its validity relies on the
choice of the particular functional form used to model the phase
evolution (a cubic polynomial in this case). We also note that the
second derivative of the spin period takes a value which is roughly one order
of magnitude smaller than that obtained considering only the first 159
days of the outburst. The best-fit solution and timing residuals are
plotted in Figure~\ref{phasecorr}.

According to this timing solution, the spin period evaluated at the
beginning of the 2009 August XMM-Newton observation (MJD 55073.61477)
is $P_{XMM}=5.7622571(2)$ s. To check this estimate, we compared this
value with that obtained by performing an epoch folding search
analysis \citep{leahy87} over the XMM-Newton data set alone,
considering 20 phase bins to sample the oscillation. The value that we
have obtained ($5.762253(8)$ s) is compatible with that predicted by
the tentative timing solution listed in Table 2, even if its larger
uncertainty with respect to the value obtained from phase fitting
makes it compatible with more than one possible timing solution.

By fitting a fifth order polynomial to the phases observed in the
interval MJD 54700.794--54940.953, \cite{Gogus10} estimated a spin
period of 5.762096529(1) at MJD 54750.0. Evaluating our solution at
that epoch yields instead an estimate of 5.76209578(4) s. The
significant difference between the two solutions is most probably due
to the different model used in the fit of the pulse phases. In
presence of a strong timing noise component, like in this case, a
higher order polynomial is able to follow more closely the variations
of the pulse phase on short time-scale, which has a significant effect
on the value of the spin period evaluated at each epoch.

In the new \XMM\, observation we find that the 0.3--12\,keV pulse
profile consists of a main peak from phases$\sim$0.3--1.0 divided into
two sub-pulses at phase$\sim$0.65, and with a flatter region between
phases$\sim$0--0.3. At lower energies, the two sub-pulses seem to
merge into one single peak as the energy increases (see
Fig.\,\ref{pfenergy}). We also computed the pulsed
fraction as $(max-min)/(max+min)$, obtaining a value of 45(6)$\%$ (see
also Tab. \ref{obslog}).  As seen in the previous 5 \XMM\,
observations, both the pulse shape and the pulsed fraction change as a
function of energy (see Fig.\,\ref{pfenergy} and
Fig.\,\ref{pulsenergy}).

\section{Spectral Analysis and Results}
\label{spectra}

To obtain the 0.5--10\,keV phase-averaged spectrum for the new \XMM\, observation we used source and background photons extracted as described in \S\ref{obs}. Then, the response matrix was created for that observation. For the present analysis we used the {\tt XSPEC} package (version 12.7.0).   

The spectrum during quiescence was well fit by an absorbed  blackbody (BB) + power law (PL) model.  Neither a single  BB  nor a PL  gave an acceptable fit. For the photoelectric absorption we used the cross-sections from \cite{balucinska92}, an the Solar abundance from \cite{anders_grevesse89}. The best-fit parameters for the absorbed  BB+PL model are $N_{\rm H}$=0.85(3)$\times10^{22}$ cm$^{-2}$, kT=0.52$\pm$0.02\,keV,  and $\Gamma$=3.87$\pm$0.13 (reduced $\chi^2$=1.06  for 165 d.o.f.).  In Table~\ref{tabspec} we  display the main spectral parameters obtained with this new \textit{XMM-Newton} observation\footnote{We note that the (BB+PL) model is unphysical, and probably tend to overpredict the (sub) keV unabsorbed flux together with the $N_{\rm H}$ value.}. We also show for comparison the results at the peak of  the 2008 outburst (more details in Rea et al. 2009).

%%%%%%%%%%%%%%%%%%%  SPECTRA     %%%%%%%%%%%%%%%%%%%%%%%%%%
%\begin{center}
\begin{figure}
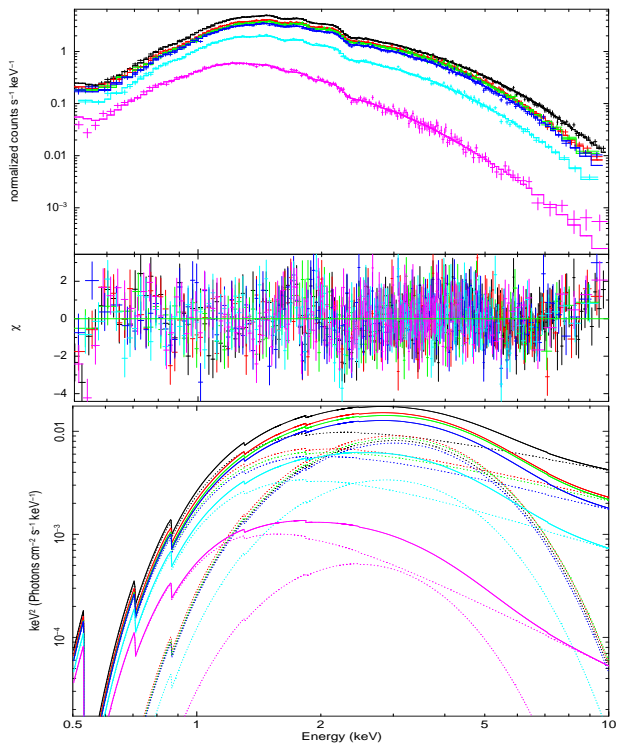

\vbox{
\psfig{figure=fig5_1.ps,width=8.021cm,height=5.25cm,angle=-90}
\hspace*{0.26cm}\psfig{figure=fig5_2.ps,width=8.0cm,height=4.5cm,angle=-90}}
\caption{Phase-averaged spectra and $\nu F_{\nu}$ plot of the fitted
  models for all the \XMM\, observations (the black, red, light green, dark blue, cyan and magenta colors correspond to increasing epoch). The continuous lines are the sum of the individual (dotted-lines) components in the bottom panel.}
\label{figspecall}
\end{figure}
%\end{center}
%%%%%%%%%%%%%%%%%%%%%%%%%%%%%%%%%%%%%%%%%%%%%%%%%%%%%%%%%%%%%%%%

%%%%%%%%%%%%%   BB+ PL    EVOL  %%%%%%%%%%%%%%%%%%%%%%%%%%%%%%%%
\begin{figure}
\psfig{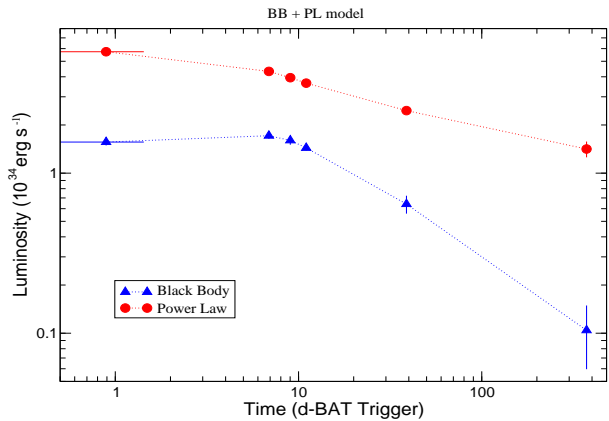}
\vspace{0.114cm}\caption{Evolution with time of the phase-averaged  BB (blue triangles) and PL (red circles) luminosities for all the \XMM\, observations for a standard BB+PL model (a distance of 2.5\,kpc was assumed).} 
\label{bb_pl_lum}
\end{figure}

%%%%%%%%%%%%%%%%%%%%%%%%%%%%%%%%%%%%%%%%%%%%%%%%%%%%%%%%%%%%%%%%

Comparing our results with the measurements reported in \cite{Gogus10}  for \textit{Swift}-XRT observations at the very end of 2009 and starting 2010, we note that their photon index ($5.04\pm0.21$) is steeper than the value found in the present study ($3.84\pm0.06$). However, in \cite{Gogus10} the \textit{Swift}-XRT spectra were extracted using observations of $\sim$12\,ks,  three times lower than the exposure time of this new \textit{XMM-Newton} observation (see Tab.~\ref{obslog}), plus the fact that the \textit{XMM-Newton} EPIC pn camera has an effective area larger than the \textit{Swift}-XRT one ($\sim\times$10). We made an attempt to reconcile these results by freezing the hydrogen column density to the value obtained by \cite{Gogus10} ($\sim$1.19$\times10^{22}$ cm$^{-2}$). We obtained a photon index of 5.59$\pm$0.13, but the reduced $\chi^2$ increased up to 1.84 (165 d.o.f.) and the residuals showed a strong sinusoidal structure.   Furthermore, we have fit the \textit{Swift}-XRT spectrum with the $N_{\rm H}$ fixed at the \textit{XMM-Newton} best-fit value.  We obtained a photon index (3.7$\pm$0.9) consistent with the \textit{XMM-Newton} one, and an acceptable fit (reduced $\chi^2$=1.1 for 78 d.o.f.). Therefore, the discrepancy may be explained due to the lower statistics of the \textit{Swift}-XRT spectra with respect to our \textit{XMM-Newton} EPIC pn observation.

In addition, the averaged spectra  showed an excess in the residuals at energies larger than 8\,keV (see  Fig.~\ref{figspecall}), that Rea et al. (2009) suggested to be due to the presence of the same hard X-ray component detected by \textit{INTEGRAL}. However, the addition of a second PL component was not significant.

%%%%%%%%%%%%%%%%%%%%   SPEC TABLE %%%%%%%%%%%%%%%%%%%%%%%%%%%%%%%%%%%%%%%%%%%%%%%%%%

\begin{table}
\begin{center}
\caption{Parameters for the spectral modelling of the phase-averaged
  spectrum of \sgrnew\, for the \XMM\ observation during quiescence. We also show for comparison the results for a \XMM\, observation  at the peak of the 2008 outburst (see also Fig.~\ref{bb_pl_lum}). The blackbody radius is calculated  assuming a distance of 2.5\,kpc (the error does not include the uncertainty in the distance).  Errors
  are at the 90\% confidence level.
  \vspace*{0.45cm}
  \label{tabspec}}
\begin{tabular}{lcc}
\hline
\hline
 \multicolumn{1}{l}{Parameters} & \multicolumn{2}{c}{Blackbody + Power-law$^a$} \\
\hline
& \multicolumn{1}{c}{2008-08-23}  & \multicolumn{1}{c}{2009-09-30} \\      
\hline
kT\,(keV) & $0.70\pm0.01$ &   $0.50\pm0.02$ \\
BB Radius (km) &  $1.41\pm0.05$ &  $0.39\pm0.05$ \\
BB flux$^b$   & $2.1\pm0.1$  & $0.14\pm0.06$ \\
 & & \\
$\Gamma$  & $2.74\pm0.02$  & $3.84\pm0.06$  \\
PL flux  &  $7.7\pm0.1$  &  $1.9\pm$0.5\\
 & & \\
Abs. Flux  &  $4.1\pm0.1$   &  $0.30\pm$0.02\\
Unab. Flux &  $9.6\pm0.1$  & $1.9\pm$0.1\\
\hline
\hline
\end{tabular}
\\\hspace*{-0.48cm}$^a$ Model: absorbed  BB +PL;  reduced $\chi^2$ (d.o.f.) = 1.15 (371);  $N_H$=$0.88(1)\times10^{22}$\,cm$^{-2}$.\\
\hspace*{-0.205cm}$^b$ Fluxes are unabsorbed in the 0.5-10\,keV range  (unless specified),
  and in units of $10^{-11}$\,erg\,cm$^{-2}$\,s$^{-1}$.

\end{center}
\end{table}

%%%%%%%%%%%%%%%%%%%%%%%%%%%%%%%%%%%%%%%%%%%%%%%%%%%%%%%%%%%%%%%%%%%%%%%%%%%%%%%% 

%%%%%%%%%%%%%%%%%%%%% PRS parameters %%%%%%%%%%%%%%%%%%%%%%%%%

\begin{figure}
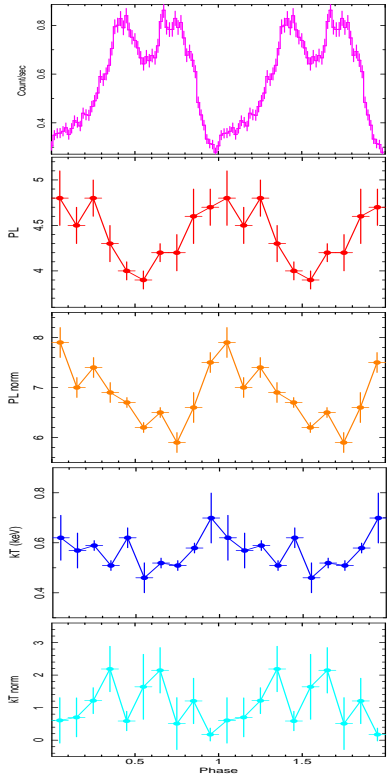

\begin{center}
\hspace{0.080cm}\vspace{0.003cm}\psfig{figure=fig7_1.ps,width=5.15cm,height=2cm,angle=-90}

\hspace{-0.015cm}\vspace{0.03cm}\psfig{figure=fig7_2.ps,width=5.25cm,height=2cm,angle=-90}

\hspace{-0.015cm}\vspace{0.03cm}\psfig{figure=fig7_3.ps,width=5.25cm,height=2cm,angle=-90}

\hspace{-0.015cm}\vspace{0.03cm}\psfig{figure=fig7_4.ps,width=5.27cm,height=2cm,angle=-90}

\hspace{-0.015cm}\psfig{figure=fig7_5.ps,width=5.25cm,height=2cm,angle=-90}

\caption{Phase-resolved spectroscopy: spectral parameters for each 0.1 phase-bin. All  spectra were fitted simultaneously with an absorbed BB + PL, keeping the $N_H$ fixed at the most accurate phase-averaged value ($N_{\rm H}=0.89(1)\times10^{22}$\cm2). The normalization of the power law is in units of 10$^{-3}$ photons keV$^{-1}$cm$^{-2}$s$^{-1}$ at 1 keV, and the blackbody normalizatio in units of $(km/kpc)^2$. }
  
\label{pps}
\end{center}
\end{figure}
%%%%%%%%%%%%%%%%%%%%%%%%%%%%%%%%%%%%%%%%%%%%%%%%%%%%%%%%%%%%%%%%

%%%%%%%%%%%%%%%%%%%%      DSP      %%%%%%%%%%%%%%%%%%%%%%%%%%%%%%%%%%%%%%%%%%%%%

\begin{figure}
\begin{center}
%\hspace*{0.005cm}
\hbox{
\vbox{
\psfig{figure=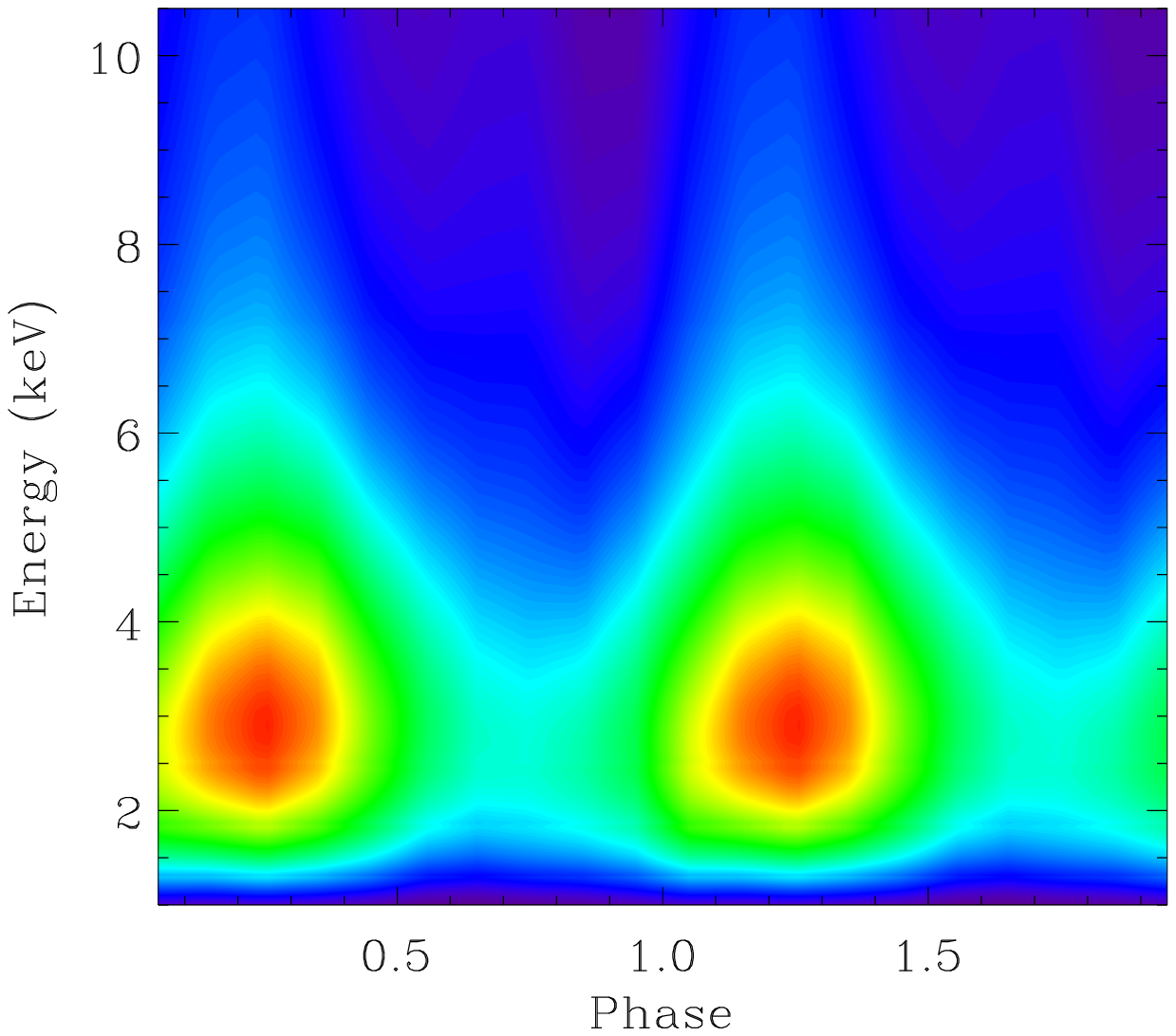,height=2.77cm,width=4cm}
\psfig{figure=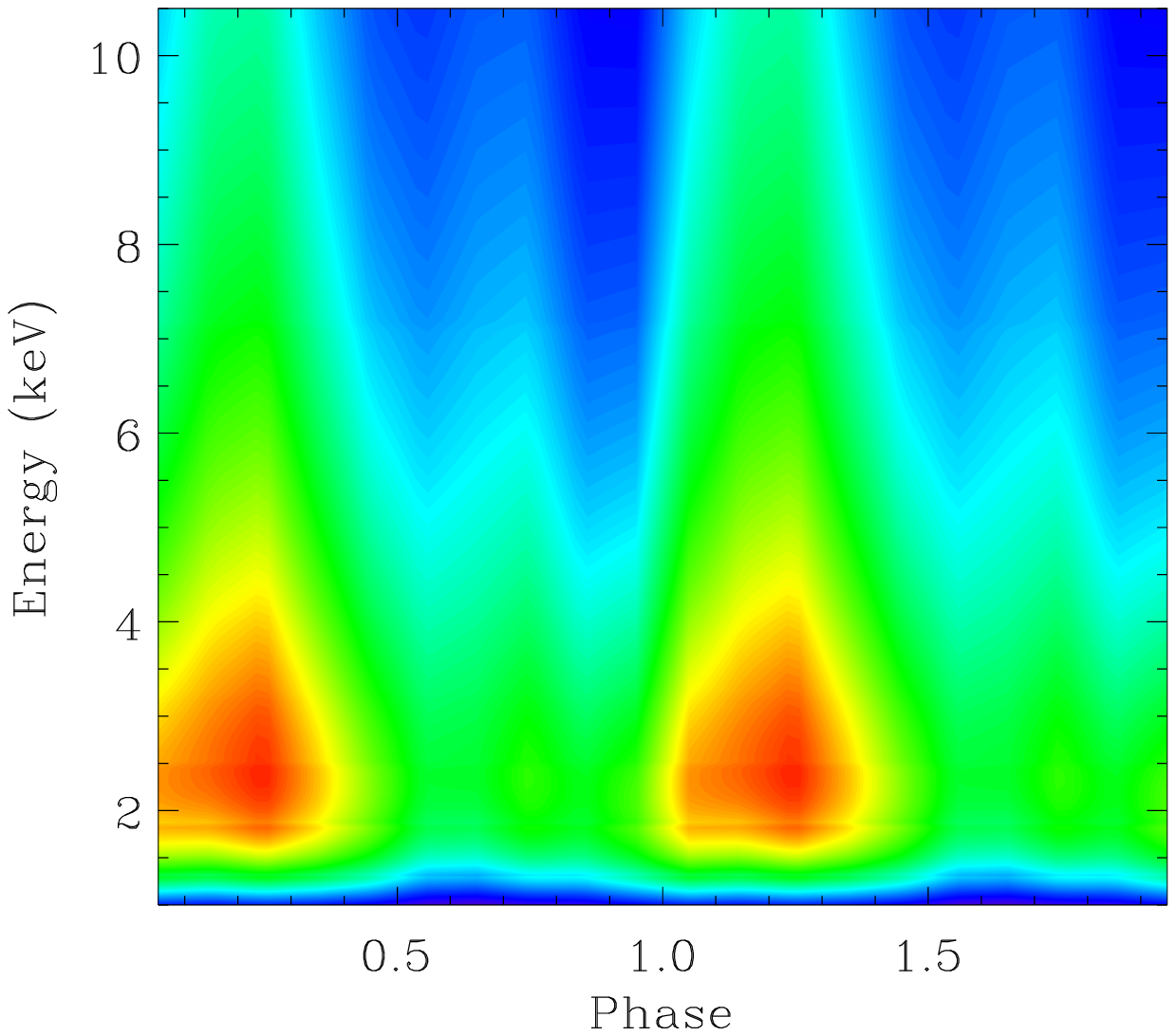,height=2.77cm,width=4cm}
\psfig{figure=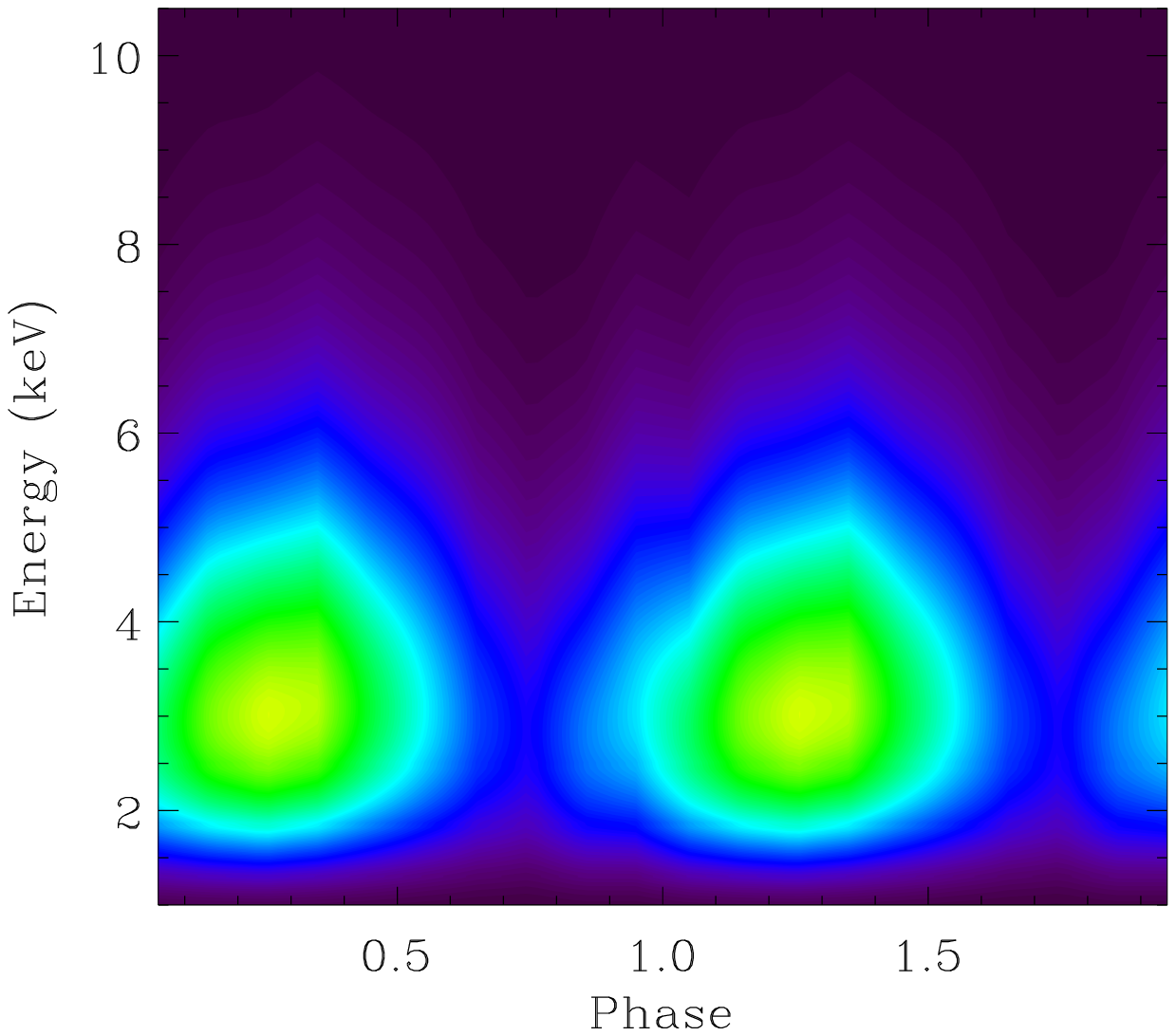,height=2.77cm,width=4cm}}
\hspace{-0.8cm}\vbox{
\psfig{figure=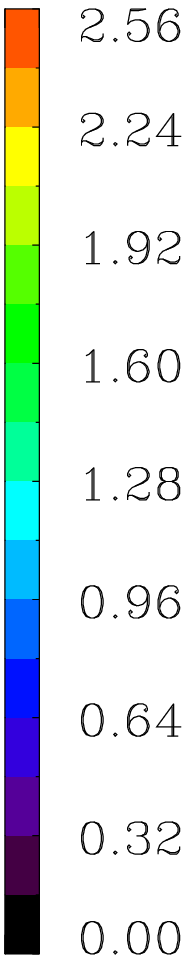,height=2.8cm,width=0.9cm}
\psfig{figure=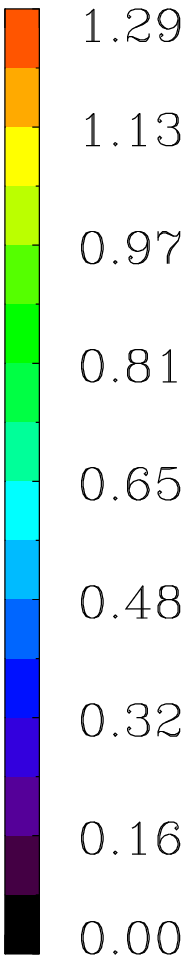,height=2.75cm,width=0.9cm}
\psfig{figure=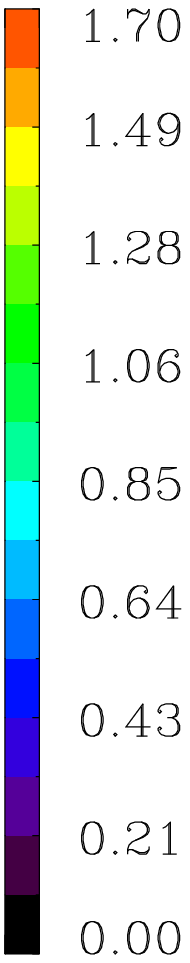,height=2.75cm,width=0.9cm}}
\vbox{
\psfig{figure=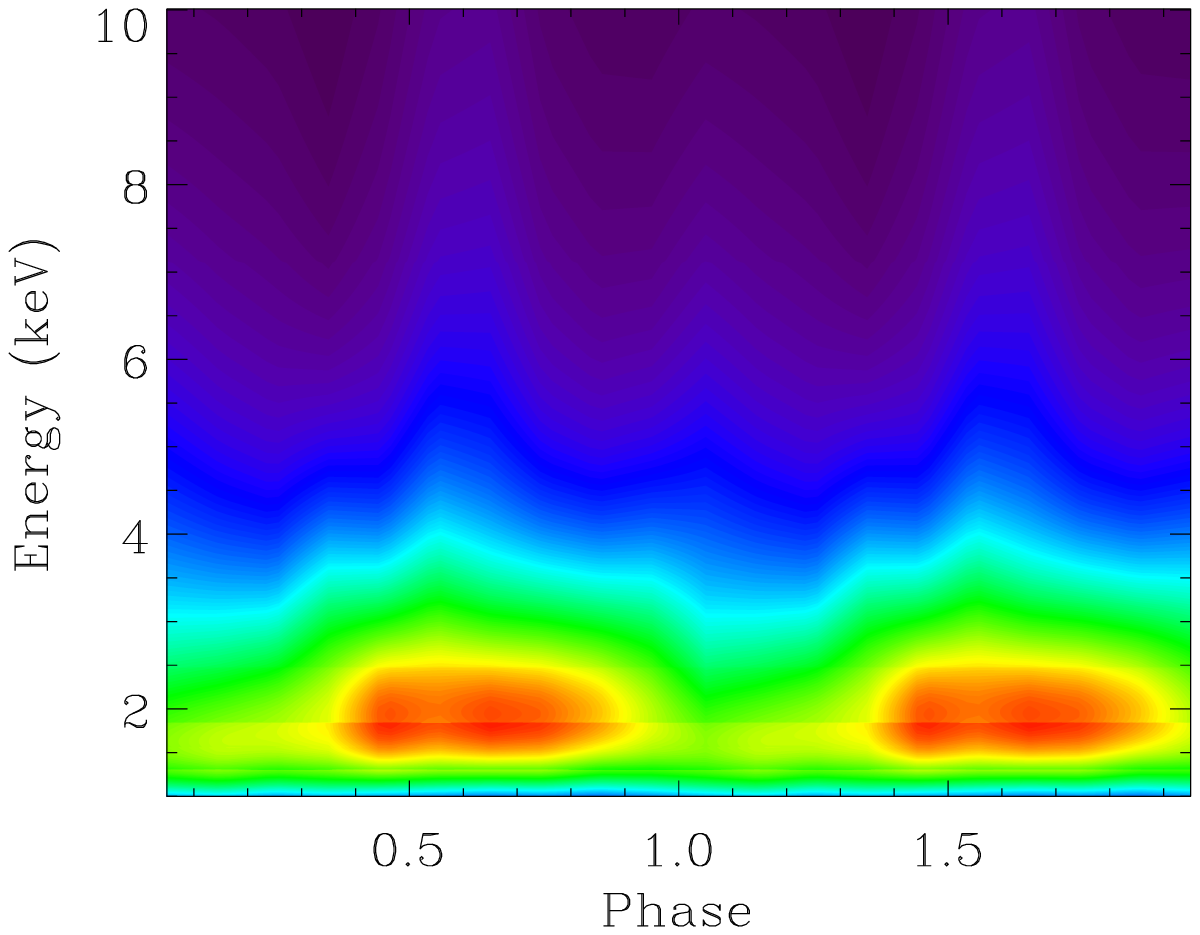,height=2.75cm,width=3cm}  %total
\psfig{figure=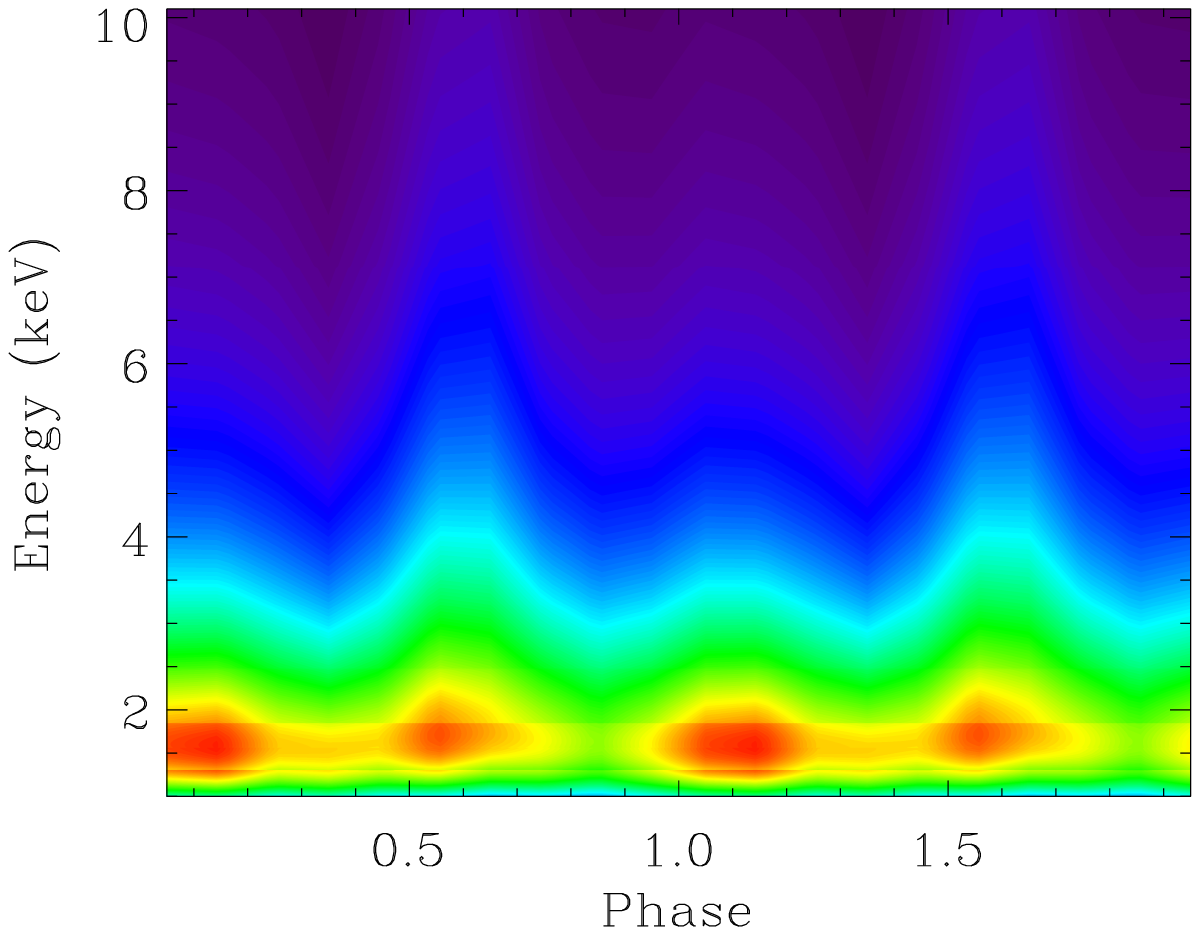,height=2.75cm,width=3cm} % pl
\psfig{figure=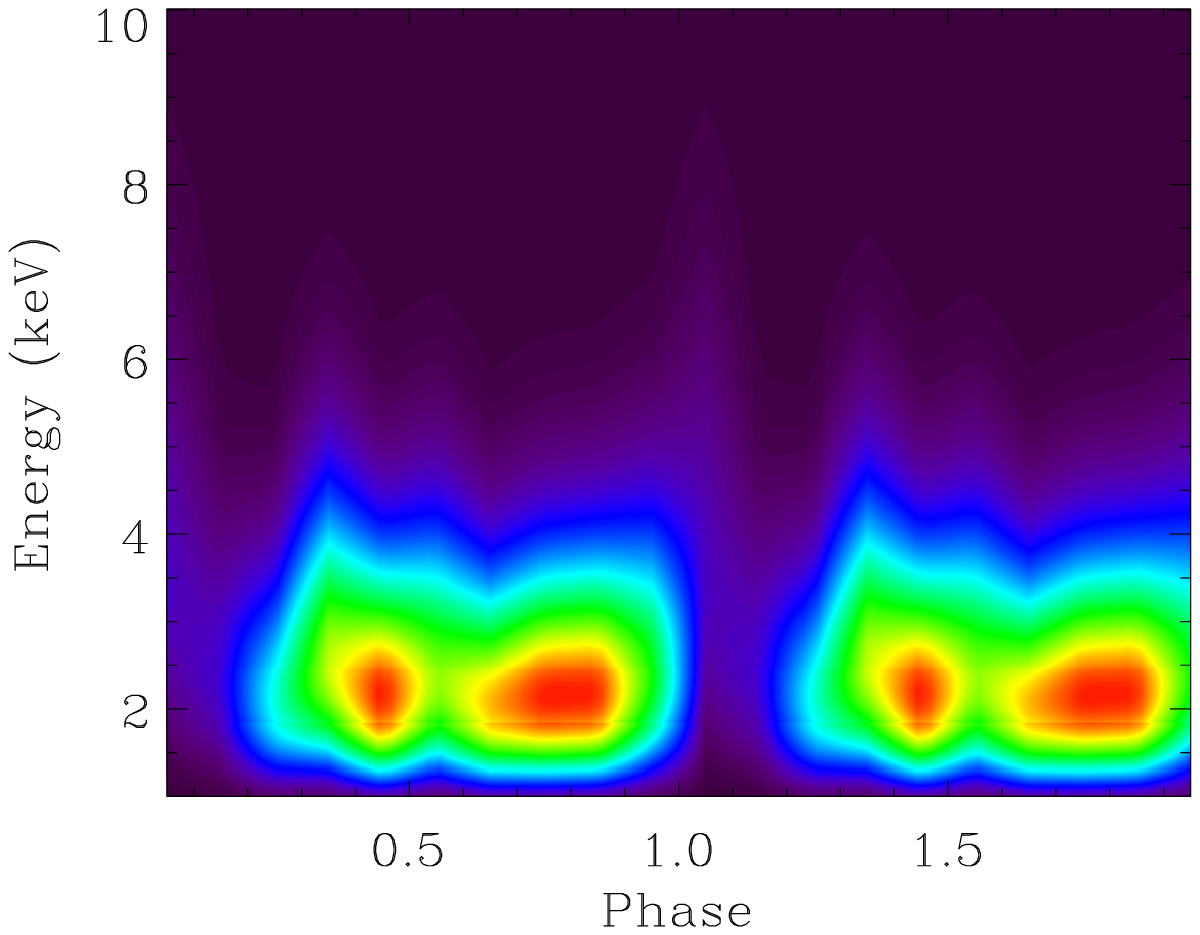,height=2.75cm,width=3cm}} %bb
\vbox{
\psfig{figure=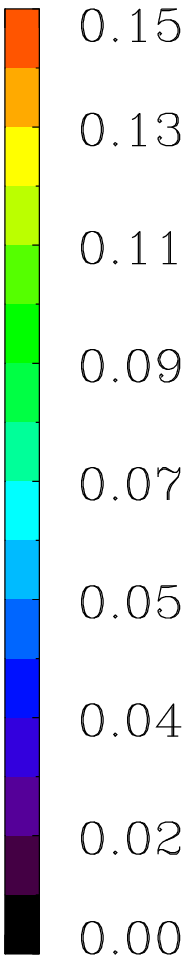,height=2.75cm,width=0.9cm}
\psfig{figure=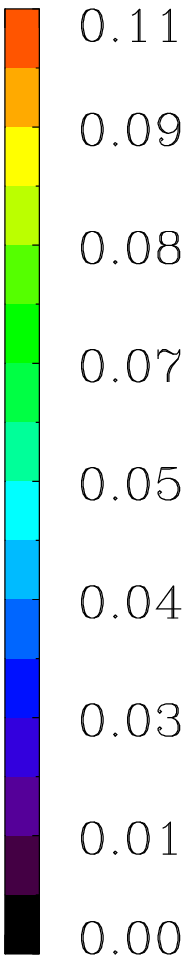,height=2.75cm,width=0.9cm}
\psfig{figure=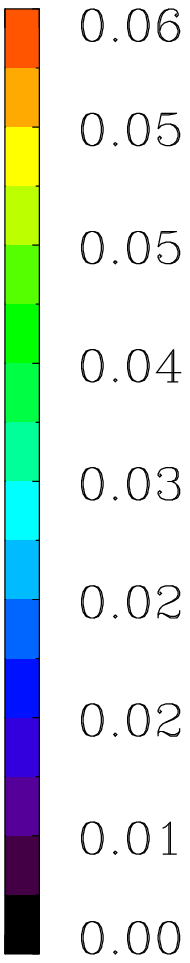,height=2.75cm,width=0.9cm}}
}
\caption{Dynamic Spectral Profiles (DSP) from two \XMM\, observations on 2008 Aug 23 (left) and 2009 Aug 30 (right). From top to bottom,  phase/energy plane the contour plots for the total (top), power-law (middle) and blackbody (bottom) $\nu$F$_\nu$ flux. The colour scale is in units of 0.01\,keV(keV\,cm$^{-2}$\,s$^{-1}$\,keV$^{-1}$).}
\label{dsp} 
\end{center}
\end{figure}

%%%%%%%%%%%%%%%%%%%%%%%%%%%%%%%%%%%%%%%%%%%%%%%%%%%%%%%%%%%%%%%%%%%%%%

To study the spectral evolution of \sgrnew\, since the 2008 Aug 22 outburst, we fitted simultaneously the spectra of all the \XMM\,  observations again with a BB plus PL model  (see  Fig.\,\ref{figspecall}), leaving all parameters free except  $N_{\rm H}$ which was constrained to be the same in all observations (final reduced $\chi^2=1.17$ for 1011 d.o.f.; see also Rea et al. 2009).  The values of the spectral parameters were not significantly different when modelling each observation separately. The measured hydrogen column density is $N_{H}=0.893(8)\times 10^{22}$\cm2 , and the absorbed flux in the 0.5--10\,keV band varied from 4.1 to 0.3$\times10^{-11}$\ergscm2\,(see also Tab.~\ref{tabspec}), corresponding to a luminosity range of 1.2 to 0.10$\times10^{35} \,d_{2.5}^2$\, \ergs\,(where $d_{2.5}$ is the source distance in units of 2.5\,kpc) (see Fig.~\ref{bb_pl_lum}).

Moreover, we have tried more sophisticated models, the RCS (Resonant Cyclotron Scattering; Rea et al. 2008b) and the NTZ  (the 3D evolution of the RCS model; Nobili et al. 2008 and Zane et al. 2009), but this resulted in poorer fits with respect to the BB+PL. %We note that the derived characteristics of the magnetosphere of these objects are still rather uncertain and need further study.

Rea et al. (2009)  found evidence that as the flux decreased the X-ray spectrum
softened during the first month after the bursting onset. In this new observation, \sgrnew\, exhibited an even softer spectrum and a flux $\sim$75 times lower than that measured during the outburst.  A similar behaviour was observed with \textit{ROSAT} back in 1992 (Rea et al. 2009). This result  is also consistent with the results obtained by \cite{Gogus10} using \textit{Swift}/XRT data.  Fitting the flux evolution in the first 548\,days after the onset of the bursting activity, we found that a double exponential function of the form  Flux(t) = $K_0 + K_1\exp{-t/t_{c1}}+K_2\exp{-t/t_{c2}}$ provided the best (although not completely satisfactorily) representation of the data (see Figure \ref{figfluxdecay}; reduced $\chi^2/d.o.f.$=5.7/51); the best values of the parameters are 
$K_0$ = 0.16$\times10^{-11}$ \ergscm2  (fixed),  
$K_1$=3.20(8)$\times10^{-11}$ \ergscm2 and 
$K_2$=0.84(8)$\times10^{-11}$ \ergscm2  , and 
$t_{c1} = 21.5(7)$ and 
$t_{c2} = 220(20)$ days.  This is similar to the outburst decays of other magnetars, usually fitted by two components (see Rea et al. 2009, and references therein).

%%%%%%%%%%%%%%%%%%%%%  Andrea  %%%%%%%%%%%%%%%%%%%%%%%%%
%\begin{center}
\begin{figure*}
\psfig{figure=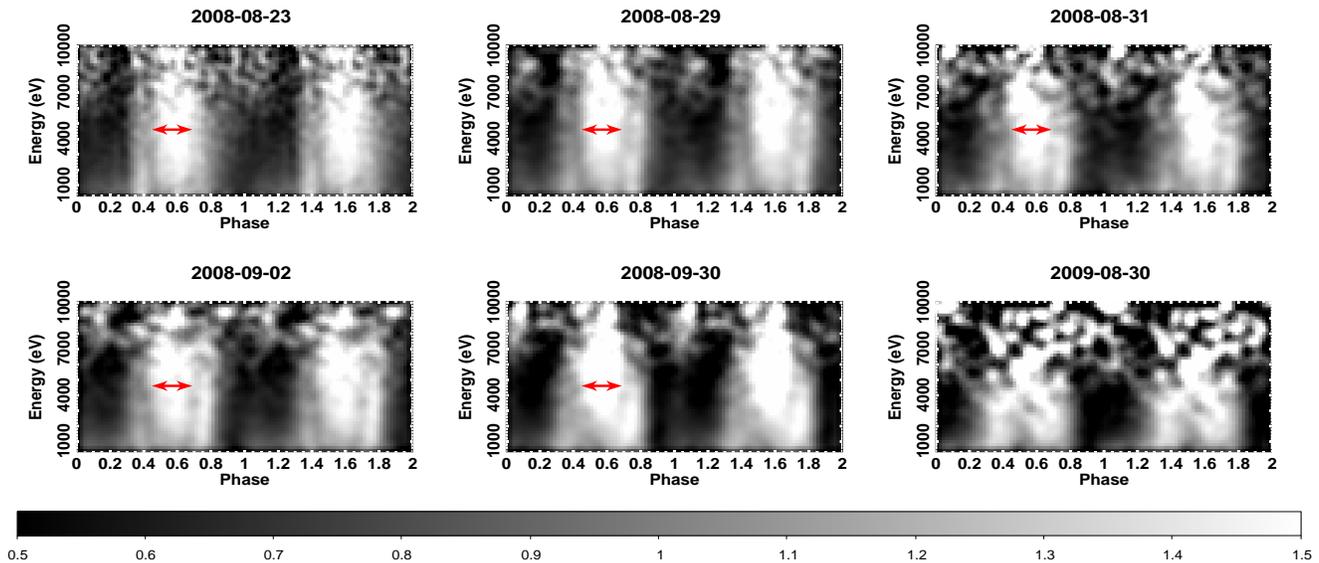,width=18cm,height=8cm}
\caption{Energy versus phase images obtained by binning the \textit{XMM-Newton}/EPIC pn source counts into 50 phase bins and energy channels of 200 eV. The double red arrow (identical for the first five observations) indicate that dark stripes, resembling the phase-variable absorption line detected in SGR\,0418+5729 (Tiengo et al. 2013), are possibly present in the first
observations.}
\label{andrea}
\end{figure*}
%\end{center}
%%%%%%%%%%%%%%%%%%%%%%%%%%%%%%%%%%%%%%%%%%%%%%%%%%%%%%%%%%%%%%%%

\subsection{Phase-resolved spectroscopy}
\label{secpps}

We performed a phase-resolved spectroscopy (PRS) for the new \xmm\,observation. We generated 10 phase-resolved spectra for this observation using the ephemeris reported in \S\,\ref{timing}.  An  absorbed BB plus PL model provided a good fit for all ten phase-resolved spectra during quiescence.  A simple absorbed BB or PL models did not yield acceptable fits. In Fig.\,\ref{pps} we have plotted the parameters derived from the PRS analysis.  In general, the spectral parameters  follow the trend seen during the outburst.  The power-law  photon index becomes softer between phases $\sim$0.0-0.4, and the blackbody temperature remained rather constant.

The pulse profiles and the spectral changes in phase and time can be globally analysed by carrying out Dynamic Spectral Profiles (DSP). In Figure~\ref{dsp} we show DSP for two \textit{XMM-Newton} observations, at the peak of the 2008 outburst (panels on the left) and during quiescence (panels on the right). Each panel show a contour plot of the $\nu$F$_\nu$ flux as a function of phase and energy, and has been derived from the 10 extracted phase-resolved spectra. The top panels correspond to the total flux, using the BB+PL model, while middle bottom panels show, respectively, the flux of the PL and BB components. The new \textit{XMM-Newton} observation during quiescence confirms the evolutionary trend of the phase-dependent spectrum during the outburst, with the PL component dominating the emission at all the times at energies above $\sim$5 keV.  However, the main component of the profiles was dominated by the BB component (see also Fig.~\ref{pps}).

To study whether the X-ray spectrum of SGR 0501+4516 has any phase-dependent absorption feature, similar to the one detected in SGR\,0418+5729 \citep{tiengo13},  we produced energy versus phase images for all the available \textit{XMM-Newton} observations by binning the source counts into 50 phase bins and
energy channels of 200 eV (and then divided by the average number of
counts in the same energy bin). The resulting images can be seen in Fig.~\ref{andrea}.  Some hints for two narrow dark stripes, indicating a possible phase-variable feature, seem to be visible at the two sides of pulse maximum in early observations. However, these deviations from the phase average spectrum are much smaller than in SGR\,0418+5729 and could not be significantly detected by fitting the corresponding phase-resolved spectra.

%%%%%%%%%%%%%%%%%%%%%%%%%%%%%%%%%%%%%%%%%%%%%%%%%%%%%%%%%%%%%%%%%

\begin{figure*}
\begin{center}
\hspace*{-0.75cm}
\hbox{
\psfig{figure=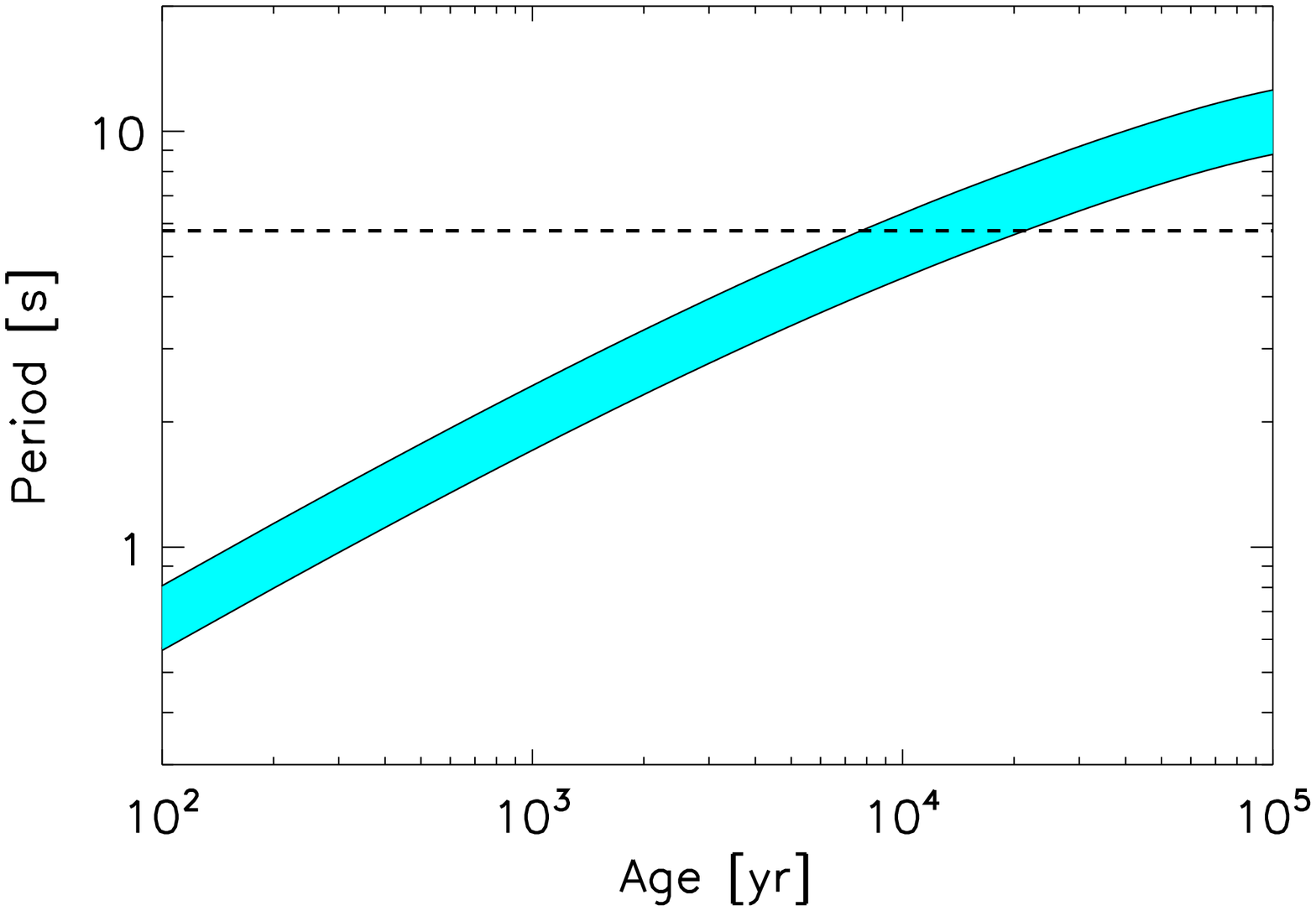,height=5.5cm,width=4.6cm}
\psfig{figure=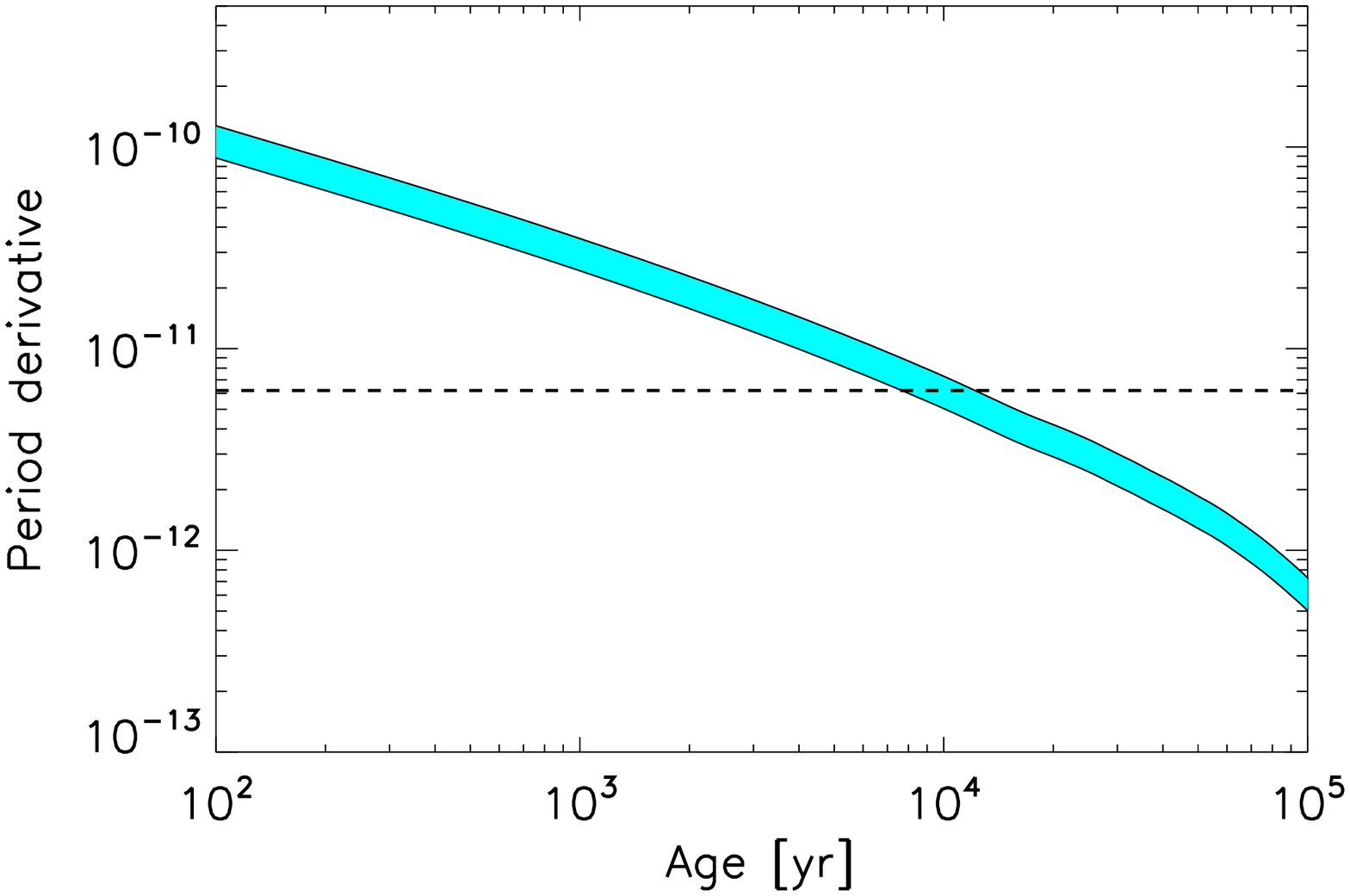,height=5.5cm,width=4.6cm}
\psfig{figure=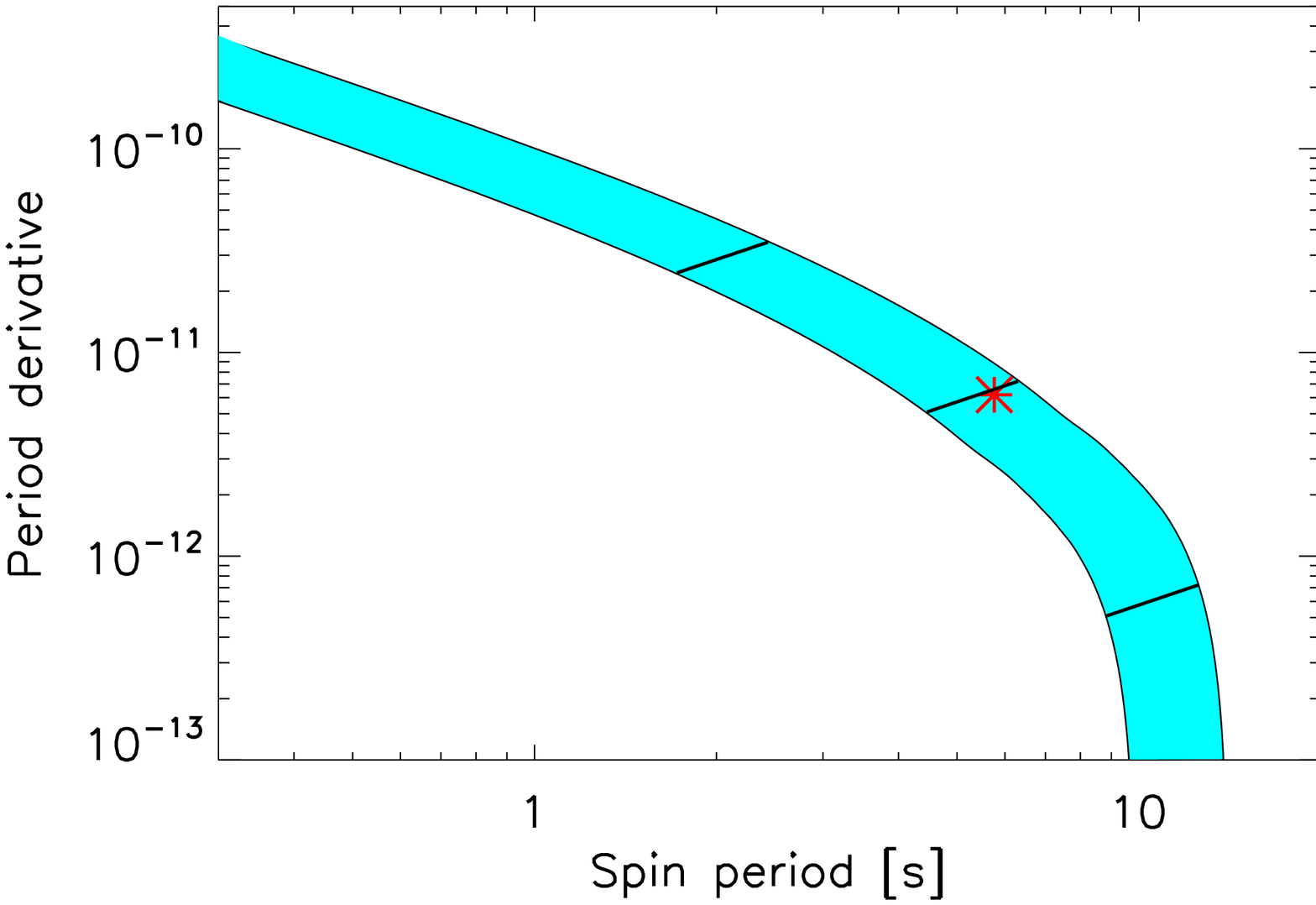,height=5.5cm,width=4.6cm}
\psfig{figure=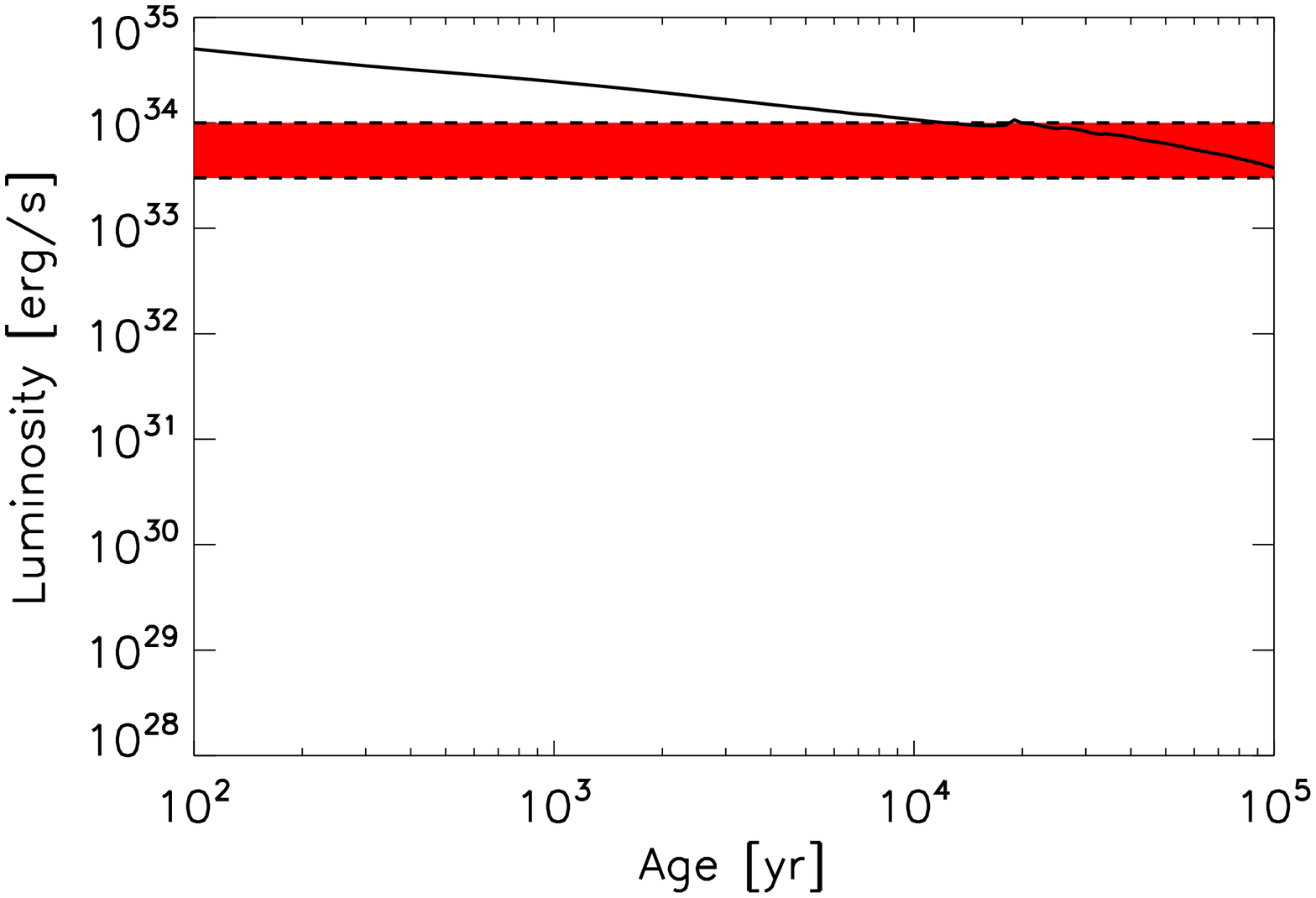,height=5.5cm,width=4.6cm}
}
\caption{Magneto-thermal evolution of a neutron star with initial magnetic field $3\times10^{14}$\,G (at the pole): from left to right, period, period derivative, evolution in the $P$-$\dot{P}$ diagram, and the thermal bolometric luminosity. The dashed lines and the asterisk mark the observed timing properties of SGR\,0501+4156. The band in the luminosity plot reflects the distance uncertainty, while the band in the timing properties corresponds to the uncertainty of the inclination angle in the theoretical model. In the third panel, the solid lines represent the real ages $t=1, 10$ and $100$ kyr.}
\label{mod} 
\end{center}
\end{figure*}

%%%%%%%%%%%%%%%%

\section{Discussion and Conclusions}
\label{discussion}

In this paper we presented an X-ray  study of the spectral and timing properties of \sgrnew during quiescence, using a new \xmm\, observation obtained on 2009 August 30. We also made use of this new observation to improve the modeling of the outburst and the spin period evolution of the source since it was first detected in 2008 Aug 22. To do this we used all the available data collected up to date from \xmm, \textit{Swift}-XRT, \textit{RXTE}, \textit{Chandra}, and \textit{Suzaku}.  
 
During quiescence both the pulse shape and the pulsed fraction change as a function of energy.  The  0.3--12\, keV pulse profile  consists mainly  of a single peak showing no very prominent features. At lower energies the pulse profile presents two peaks that seem to get dissolved into one as the energy increases. Moreover, the pulsed fraction increases at any given energy range as the outburst evolves (see Fig.~3).

The  spectrum was well fit by an absorbed  BB + PL model.  The best-fit parameters for this  model are $N_H$=0.85(3)$\times10^{22}$\,cm$^{-2}$, kT=0.52$\pm$0.02\,keV,  and $\Gamma$=3.87$\pm$0.13.  The source shows an absorbed  flux $\sim$75 times lower than that measured during the 2008 outburst, and a rather soft  spectrum, with the same  value of the blackbody temperature  observed with \textit{ROSAT} back in 1992 and with \textit{Swift}-XRT in 2009-2010. The 0.1--2.4 keV observed flux was F$_X \sim$1.8$\times$10$^{-12}$\,erg cm$^{-2}$ s$^{-1}$, in line with that observed by ROSAT before the outburst.

All magnetar outbursts detected so far were characterized by the presence of a hot spot, which cools down until quiescence is reached; the peak luminosity is of the order of a $\times10^{35-36}$\ergs \citep[see][]{pons_rea12}. Magnetar outbursts are believed to be due to a sudden crustal crack consequence of their unstable magnetic field configuration, when the twisted magnetic field in its constant movement toward a stable/quiet assessment, stresses the crust until its breaking point. The large amount of energy that is released in this event is expected to be of the order of $\sim10^{42-44}$ ergs. Part of this energy will heat the crust (explaining the sudden increase of the surface temperature soon after the outburst onset), part of it will be spent in accelerating particles in the magnetosphere (explaining the hardening of the spectral shape), and a large fraction will be released in neutrinos. The limiting luminosity is a consequence of the neutrino emission that regulates the release of most of the energy dumped in the crust as soon as a temperature reaches here a value of 4$\times$10$^9$ K \citep{pons_rea12}. The total energy released by \sgrnew's outburst in the 0.5--10\,keV band  is $\sim2\times10^{41}$ ergs, in line with what expected for a typical large scale crustal event. On the other hand, all typical outburst characteristics were observed for \sgrnew\, (see \S\ref{spectra}), as well as a clear dependence on the rotational phase of the crustal heating, which was localized in a few degrees of surface, creating a localized hot spot, possibly on the magnetic pole, which cooled down until its quiescent emission level.

The long-term magneto-thermal evolution, simulated with the \cite{vigano12} code, indicates that this source is consistent with being born $\sim10$\,kyr ago (not too far from its 16\,kyr characteristic age) with an initial dipolar magnetic field of $3\times10^{14}$ G.  This exact value depends, by a factor $<$2, on the parameters of the neutron star (mass, radius, inclination angle) and on the poorly known internal magnetic configuration. In Figure ~\ref{mod}, we show the evolution of period, period derivative, the source track in the $P$-$\dot{P}$ diagram, and the thermal, bolometric luminosity.

As generally found in magnetars, the temperature inferred from the BB+PL fit is of the same order of, although somewhat higher than, what predicted by the evolutionary model (100-200 eV, depending on the details and the internal geometry). The mismatch (by a factor $\sim 2$--3) is well within what can be expected from the assumptions of both the emission and evolution models. It is known that spectral fitting with resonant Compton scattering models results in a lower temperature, closer to that of the cooling models, even keeping the assumption of blackbody seed photons. Actually, thermal radiation from the star surface is likely different from a simple blackbody, because of local reprocessing by some sort of atmosphere or, on the contrary, because the surface is in a condensed state. Moreover, the hot spot temperature is not uniform and geometrical effects (viewing angle, inclination of the magnetic axis) come into play; these are not accounted for as yet. Given these uncertainties, we rather compare the luminosity, which depends on the distance, but not much on the details of the model.

%%%%%%%%%%%%%%%%%%%%%%%%%%%%%%%%%%%%%%%%%%%%%%%%%%%%%%%%%%%%%%%%%

\begin{figure}
\begin{center}
\psfig{figure=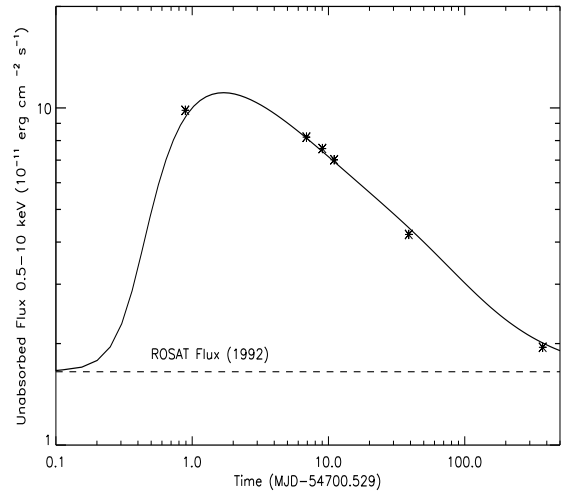,height=7cm,width=8cm}
\caption{Outburst model from Pons and Rea (2012) superimposed to the unabsorbed 0.5--10\,keV flux decay of SGR\,0501+4156  using all the \textit{XMM-Newton} observations (see the text for details).}
\label{mod_jose} 
\end{center}
\end{figure}

%%%%%%%%%%%%%%%%

We note that this source belong to a vast group of magnetars  (1E 1547-5408, SGR 1627-41, SGR 1745-2900, XTE J1810-197, SGR 1833-0832, 4U 0142+614, CXO 1647-4552), which timing properties are linked by the same evolutionary track, compatible with being born with an initial field of $2-3\times 10^{14}$ G, as discussed in \cite{vigano13}. In this scenario, most magnetars are born with a narrow range of magnetic field values, and the timing and spectral differences between them are due only to their particular ages.

We have also compared the observed outburst decay with the detailed theoretical model presented in \cite{pons_rea12}, and used to reproduce the outburst decay of other magnetars \citep{rea12,rea13}. Starting with a neutron star with $B=2\times10^{14}$ G and the crust temperature fixed to reproduce the flux observed by ROSAT in 1992 at the estimated distance, we simulate the starquake by injecting $1.5 \times 10^{42}$ erg in the crust layer between $7 \times 10^8$ and $4 \times 10^{10}$ g cm$^{-3}$ in a polar cap of about 30 degrees around the pole. We then follow the thermal evolution
of the crust until it returns to the original state. The results are shown in Figure~\ref{mod_jose}, where we compare the theoretical model with the \textit{XMM-Newton} observations.

\section*{Acknowledgements}

This work was supported by the grants AYA2012-39303, SGR2009-811,  and iLINK2011-0303. AP is supported by a Juan de la Cierva Fellowship in IEEC. NR is supported by a Ramon y Cajal fellowship and by an NWO Vidi Award. DV was supported by the grants AYA2010-21097-C03-02, ACOMP/2012/135, AYA 2012-39303 and SGR 2009-811.

%\bibliographystyle{astron}
%\bibliography{sgr0501_quiesc}

\end{document}